\documentclass[sigconf,natbib=true,screen=true]{acmart}

\copyrightyear{2026}
\acmYear{2026}
\setcopyright{cc}
\setcctype{by}
\acmConference[SIGIR '26]{Proceedings of the 49th International ACM SIGIR Conference on Research and Development in Information Retrieval}{July 20--24, 2026}{Melbourne, VIC, Australia}
\acmBooktitle{Proceedings of the 49th International ACM SIGIR Conference on Research and Development in Information Retrieval (SIGIR '26), July 20--24, 2026, Melbourne, VIC, Australia}
\acmDOI{10.1145/3805712.3808557}
\acmISBN{979-8-4007-2599-9/2026/07}
\usepackage{acronym}
\usepackage[skip=0pt]{caption}
\usepackage[inline]{enumitem}
\usepackage{enumitem}
\usepackage{booktabs} % For formal tables
\usepackage{multirow}
\usepackage{array}
\usepackage{pifont}
\usepackage{xspace}
\usepackage{subcaption}
\usepackage{makecell}
\usepackage{CJKutf8}
\usepackage{tcolorbox}
\usepackage[para]{footmisc}
\usepackage{amsfonts}
\usepackage{rotating}
\usepackage[dvipsnames]{xcolor}
\usepackage{todonotes}
\usepackage{tabularx}
\acrodef{QF-DC}{query filtering mechanism based on document relevance and conversation alignment}

\acrodef{RAG}{retrieval-augmented generation}

\acrodef{RCD}{Retrieval from Conversational Dialogues}
\acrodef{PSC}{proactive search in conversations}
\acrodef{CPS}{conversational proactive search}
\acrodef{PS}{proactive search}

\acrodef{GR}{generative retrieval}
\acrodef{LLM}{large language model}
\acrodef{QPP}{query performance prediction}
\acrodef{IR}{information retrieval}
\acrodef{NLP}{natural language processing}
\acrodef{PEFT}{parameter-efficient fine-tuning}
\acrodef{ICL}{in-context learning}
\acrodef{LoRA}{low-rank adaptation}

\acrodef{RR}{reciprocal rank}
\acrodef{AP}{Average Precision}
\acrodef{nDCG}{normalized discounted cumulative gain}

\acrodef{HSD}{Tukey's honestly significant difference}
\acrodef{ANOVA}{analysis of variance}

\acrodef{UEF}{utility estimation framework}
\acrodef{QPP-PRP}{pairwise rank preference-based QPP}
\acrodef{M-QPPF}{multi-task query performance prediction framework}
\acrodef{WRIG}{weighted relative information gain-based model}
\acrodef{NLP}{natural language processing}
\acrodef{CIS}{conversational information seeking}
\acrodef{ANCE}{Approximate nearest neighbor Negative Contrastive Estimation}

\acrodef{RL}{reinforcement learning}

\newcommand\setItemnumber[1]{\setcounter{enumi}{\numexpr#1-1\relax}}

\newcommand{\header}[1]{\vspace*{1mm}\noindent\textbf{#1}.}



\author{Chuan Meng}
\orcid{0000-0002-1434-7596}
\affiliation{%
  \institution{The University of Edinburgh}
  \city{Edinburgh}
  \country{United Kingdom}
}
\email{chuan.meng@ed.ac.uk}

\author{Litu Ou}
\orcid{0009-0005-8380-1368}
\affiliation{%
  \institution{The University of Edinburgh}
  \city{Edinburgh}
  \country{United Kingdom}
}
\email{litu.ou@ed.ac.uk}

\author{Sean MacAvaney}
\orcid{0000-0002-8914-2659}
\affiliation{%
  \institution{University of Glasgow}
  \city{Glasgow}
  \country{United Kingdom}
}
\email{sean.macavaney@glasgow.ac.uk}

\author{Jeff Dalton}
\orcid{0000-0003-2422-8651}
\affiliation{%
  \institution{The University of Edinburgh}
  \city{Edinburgh}
  \country{United Kingdom}
}
\email{jeff.dalton@ed.ac.uk}

\begin{document}

%\title[Text Ranking in Deep Research]{Text Ranking in Deep Research}
%\title[Revisiting Information Retrieval in Deep Research]{Revisiting Information Retrieval in Deep Research}
\title[Revisiting Text Ranking in Deep Research]{Revisiting Text Ranking in Deep Research}

\renewcommand{\shortauthors}{Chuan Meng et al.}

% align the the types of queries to to what the retrievers expect.

\begin{abstract}
Deep research has emerged as an important task that aims to address hard queries that need extensive open-web exploration.
To tackle it, most prior work equips large language model (LLM)-based agents with opaque web search APIs, enabling agents to iteratively issue search queries, retrieve external evidence, and reason over it.
Despite search's essential role in deep research, black-box web search APIs leave the behaviour of established text ranking methods in deep research largely unclear. 
% hinder systematic analysis of search components, 
%
To fill this gap, we reproduce key findings and best practices for text ranking methods in deep research.
We examine their effectiveness from three perspectives:
\begin{enumerate*}[label=(\roman*)]
\item retrieval units (documents vs.\ passages),
\item pipeline configurations (different retrievers, re-rankers, and re-ranking depths), and
\item query characteristics (the mismatch between agent-issued queries and the training queries of text rankers).
\end{enumerate*}
We perform experiments on BrowseComp-Plus, a deep research dataset with a fixed corpus, evaluating 2 open-source agents, 5 retrievers, and 3 re-rankers.
We find that agent-issued queries typically follow web-search-style syntax (e.g., quoted exact
matches), favouring lexical, learned sparse, and multi-vector retrievers; passage-level units are more efficient under limited context windows, and avoid the difficulties of document length normalisation in lexical retrieval; re-ranking is highly effective.
We further propose a query-to-question (Q2Q) method that translates agent-issued queries into natural-language questions, significantly reducing the query mismatch.
\end{abstract}

\maketitle
\acresetall

%
%2.2 Simulate different levels of relevance, impact on accuracy

\section{Introduction}
Text ranking is a core part of information retrieval (IR)~\citep{Robertson1994OkapiAT,lassance2024splade,ma2024fine,zhang2025qwen3,santhanam2022colbertv2,nogueira2020document,weller2025rank}.
It aims to produce an ordered list of texts retrieved from a corpus in response to a query~\citep{lin2022pretrained}.
%
%Text ranking methods have been studied across diverse settings, from single-hop~\citep{bajaj2018ms,kwiatkowski2019natural} and multi-hop search~\citep{ho2020constructing,yang2018hotpotqa} to reasoning-intensive search~\citep{shao2025reasonir,hongjinbright} and retrieval-augmented generation (RAG)~\citep{su2025parametric,mo2025uniconv,asai2024selfrag}.
%
Recently, the IR community has witnessed the emergence of \textit{deep research}~\citep{shi2025deep} scenarios.
Deep research aims to answer multi-hop, reasoning-intensive queries that require extensive exploration of the open web and are difficult for both humans and large language models (LLMs)~\citep{wei2025browsecomp}.
To tackle this task, a growing number of studies build agents that interact with live web search APIs to obtain information~~\citep{zhou2025browsecomp,li2025websailor,liu2025webexplorer}.
Such agents typically use LLMs as their decision-making core and perform multiple rounds of chain-of-thought (CoT) reasoning~\citep{weiemergent} and search invocations~\citep{jin2025searchr,song2025r1,li2025search}.

%Text ranking methods have been widely studied across a range of scenarios, including dealing with single-hop~\citep{bajaj2018ms,kwiatkowski2019natural} and multi-hop~\citep{ho2020constructing,yang2018hotpotqa}, and reasoning-intensive queries~\citep{hongjinbright}, as well as applications in retrieval-augmented generation (RAG)~\citep{su2025parametric,mo2025uniconv,asai2024selfrag}.
%

% have shown that building LLM-based agents that access web search tools can effectively address deep research tasks~\citep{zhou2025browsecomp,hu2026sage,li2025websailor,liu2025webexplorer}.
% Deep research is an important task because it can significantly reduce the time required for human experts to answer challenging questions~\citep{java2025characterizing}.
% Deep research is important because agents that can perform this task well can substantially reduce the time human experts need to answer challenging questions~\citep{java2025characterizing}.
%\footnote{For instance, deep research queries in the BrowseComp dataset can be so challenging that LLMs achieve near-zero accuracy when relying solely on parametric knowledge, and humans may be unable to solve them within two hours of searching~\citep{wei2025browsecomp}.
%

\header{Motivation}
Despite the essential role of search in deep research, how existing text ranking methods perform in this setting remains poorly understood.
Most prior studies rely on black-box live web search APIs~\citep{xu2026self,zhou2025browsecomp}, which lack transparency, thereby hindering systematic analysis and a clear understanding of the contribution of search components~\citep{chen2025browsecomp,hu2026sage}.
%
%Although recent work~\citep{chen2025browsecomp} builds a deep research dataset with a fixed document corpus and human-verified relevance judgments, it evaluates only two retrievers on the dataset.
%Very recent work~\citep{sharifymoghaddam2026rerank} focuses on exploring only one specific type of re-ranker in this context.
%
%Therefore, the overall picture of the role of existing text ranking methods in the context of deep research remains unclear.

\header{Research goal}
In this paper, we examine \textit{to what extent established findings and best practices in text ranking methods are generalisable to the deep research setup}.
%
%Following the reproducibility track's objectives, this study is conducted by a \textit{different team} under a \textit{new experimental setup} (i.e., exploring text ranking methods in the context of deep research), aligned with the ACM's (v1.1) definition of a \textit{replication} study.
%
%We revisit the following findings from the literature on text ranking:
%\todo{I would expect these to be aligned with the RQs, research gaps, and the list in the abstract.}
%
%\begin{enumerate*}[label=(\roman*)]
%    \item neural retrievers often underperform traditional lexical sparse retrievers such as BM25~\citep{Robertson1994OkapiAT} in scenarios on which they have not been trained~\citep{thakur2021beir};
%    \item re-ranking has been shown to effectively improve ranking performance~\citep{ma2024fine,nogueira2019passage}, and deeper re-ranking depth generally leads to better re-ranking performance in most cases~\citep{meng2024ranked}; and
%     \item reasoning-based re-rankers outperform their non-reasoning counterparts and demonstrate greater robustness~\citep{weller2025rank,yang2025rank}.
%\end{enumerate*}
%\header{Research gaps}
We identify three research gaps in text ranking for deep research that have received limited attention.

First, \textit{passage-level information units have received little attention in deep research}.
Most existing studies build deep research agents that search and read at the document level (i.e., full web pages).
Because feeding full web pages to LLM-based agents quickly exhausts the context window, prior work~\citep{sharifymoghaddam2026rerank,chen2025browsecomp} typically returns truncated documents to agents, which may remove relevant content and lead to information loss.
Although prior work introduces an additional full-document read tool that agents can invoke to access complete documents~\citep{chen2025browsecomp}, it adds system complexity.
There is strong motivation to explore passage-level units in deep research:
\begin{enumerate*}[label=(\roman*)]
%\item their concise nature is more efficient for an LLM's token budget, making them more efficient under limited context windows;
\item their concise nature makes them more efficient under limited context-window budgets;
\item they allow agents to access any relevant segments within a document, avoiding information loss from document truncation;
\item passages can enhance lexical retrieval by avoiding the difficulties of document length normalisation~\citep{kaszkiel1997passage}; and
\item a large body of neural retrievers has been developed for passage retrieval~\citep{lassance2024splade,santhanam2022colbertv2,formal2021spladev2,formal2021splade}, but it remains unclear how well they perform in deep research.
%Passages enables agents to access any segment within a document and friendly to LLM's token budget.
%passages are more convenient for humans to read than long documents~\citep{kaszkiel1997passage}, which may also be more efficient for LLMs to process.  %benefit LLM-based agents
\end{enumerate*}
To address this gap, we ask \textbf{RQ1:} \textit{To what extent are existing retrievers effective in deep research under passage-level and document-level retrieval units?}
In this RQ, we revisit key findings that neural retrievers often underperform lexical methods (e.g., BM25~\citep{Robertson1994OkapiAT}) on out-of-domain data~\citep{thakur2021beir}, and that learned sparse and multi-vector dense (a.k.a. late-interaction) retrievers generalise better than single-vector dense retrievers on out-of-domain data~\citep{formal2021spladev2,thakur2021beir}.
%, and that passages lessen document length normalisation effects~\citep{kaszkiel1997passage}.

Second, \textit{our understanding of re-ranking in deep research remains limited}.
Re-ranking plays an important role in lifting relevant documents to the top of ranked lists in traditional search settings~\citep{nogueira2019passage}.
It remains unclear whether widely-used re-ranking methods~\citep{weller2025rank,ma2024fine,nogueira2020document} provide consistent benefits when the content consumer is an LLM-based agent rather than a human user.
Despite recent work~\citep{sharifymoghaddam2026rerank} examining one specific re-ranking method with a single retriever, a systematic evaluation across various ranking configurations is still lacking in deep research.
To address this gap, we ask \textbf{RQ2:} \textit{To what extent is a re-ranking stage effective in deep research under different initial retrievers, re-ranker types, and re-ranking cut-offs?}
In this RQ, we revisit established findings that re-ranking effectively improves ranking performance~\citep{ma2024fine,thakur2021beir,nogueira2019passage}, deeper re-ranking generally produces better results~\citep{meng2024ranked}, and reasoning-based re-rankers outperform their non-reasoning counterparts~\citep{weller2025rank,yang2025rank}.

Third, \textit{the potential mismatch between agent-issued queries and the queries used to train existing text ranking methods remains underexplored}.
Many text ranking methods in the IR community are trained on natural-language-style questions, such as those in MS MARCO~\citep{bajaj2018ms}.
However, agent-issued queries may not align with the queries these methods expect.
Prior studies show that a query-format mismatch between training and inference can significantly hurt neural retrieval quality~\citep{meng2025bridging,zhuang2022bridging}.
It remains unclear how such a potential mismatch affects the performance of existing text ranking methods in deep research. 
To address this gap, we ask \textbf{RQ3:} \textit{To what extent does the mismatch between agent-issued queries and the training queries used for text ranking methods affect their performance?}

\header{Experiments}
We perform experiments on BrowseComp-Plus~\citep{chen2025browsecomp}, a deep research dataset that provides a fixed document corpus and human-verified relevance judgments.
%\footnote{To the best of our knowledge, at the time of writing, this is the only publicly available deep research dataset providing this setup.}
%
To ensure broad coverage, we use widely-used retrievers spanning 4 main paradigms in modern IR: lexical-based sparse (BM25~\citep{Robertson1994OkapiAT}), learned sparse (SPLADE-v3~\citep{lassance2024splade}), single-vector dense (RepLLaMA~\citep{ma2024fine}, Qwen3-Embed~\citep{zhang2025qwen3}), and multi-vector dense retrievers (ColBERTv2~\citep{santhanam2022colbertv2}).
For re-ranking, we select methods that trade off effectiveness and efficiency at 3 operational points: a relatively inexpensive re-ranker (monoT5-3B~\citep{nogueira2020document}), an LLM-based re-ranker (RankLLaMA-7B~\citep{ma2024fine}), and a CoT-based reasoning re-ranker (Rank1-7B~\citep{weller2025rank}), which generates additional reasoning tokens.
%
%For agents that access these ranking methods, 
We use two open-source LLM-based agents: gpt-oss-20b~\citep{agarwal2025gpt} and GLM-4.7-Flash (30B)~\citep{zeng2025glm}.

%For re-ranking, we select methods reflecting the evolution of neural re-ranking in IR: a re-ranker based on an encoder–decoder pretrained model (monoT5~\citep{nogueira2020document}), an LLM-based re-ranker (RankLLaMA~\citep{ma2024fine}), and a LLM-based re-ranker using CoT-based reasoning (Rank1~\citep{weller2025rank}).

\header{Findings}
For RQ1, our findings are fivefold:
\begin{enumerate*}[label=(\roman*)]
\item 
%Retrievers operating on the passage corpus achieve higher answer accuracy than those using the document corpus, although retrieval performance is typically higher on the document corpus.
The concise nature of passage-level units enables more search and reasoning iterations before reaching context-window limits, resulting in higher answer accuracy than document-level units (without a full-document reader), particularly for the gpt-oss-20b agent that has shorter context windows.
\item BM25 on the passage corpus outperforms neural retrievers in most cases (gpt-oss-20b with BM25 achieves the highest accuracy of 0.572 across all retrieval settings in our study).
We find agent-issued queries tend to follow a \textit{web-search style with keywords, phrases, and quotation marks for exact matching} (see Table~\ref{table_query_example}), favouring lexical retrievers such as BM25.
\item On the document corpus, BM25 performs worst under the parameter settings used in prior work~\citep{chen2025browsecomp}. 
We find that these parameters lack proper document-length normalisation. 
With document-oriented parameters, however, BM25 becomes highly competitive. 
%This highlights the sensitivity of lexical retrievers to length normalisation and shows the advantage of passage-level units, which reduce reliance on document-length normalisation.
This highlights BM25’s sensitivity to length normalisation and the advantage of passage-level units, which reduce reliance on document-length normalisation.
\item learned sparse~\citep{lassance2024splade} and multi-vector dense~\citep{santhanam2022colbertv2} retrievers with only millions of parameters generalise better to web-search-style queries than 7B/8B single-vector dense models.
\item Enabling a full-document reader on truncated documents generally improves answer accuracy and reduces search calls, suggesting that the reader complements truncated inputs.
In contrast, adding the reader to the passage corpus slightly degrades performance, likely because passage retrieval already provides access to any segments within a document, rendering the reader redundant.
%In contrast, adding the same tool slightly degrades performance on the passage corpus. 
%We hypothesise this is because passage retrieval already enables agents to get any relevant segment within a document; therefore, a full-document reader might introduce redundant functionality.
% and provide little value.
\end{enumerate*}
%Augmenting truncated documents with a full-document reader tool generally improves answer accuracy and reduces the number of search calls, indicating that the reader tool complements truncated documents.

For RQ2, re-ranking consistently improves ranking effectiveness and answer accuracy while reducing search calls, confirming its important role in deep research.
These gains are further amplified by deeper re-ranking depths.
%and stronger initial retrievers.
%Deeper re-ranking depths and stronger initial retrievers generally yield higher effectiveness and fewer search calls.
%
Notably, the BM25--monoT5-3B pipeline with gpt-oss-20b achieves the best results in our work, reaching 0.716 recall and 0.689 accuracy. 
Despite using only a 20B agent with BM25 and a 3B re-ranker, this setup approaches the 0.701 accuracy of a GPT-5–based agent (Table~1 in~\citep{chen2025browsecomp}).
The reasoning-based Rank1~\citep{weller2025rank} shows no clear advantage over non-reasoning methods, as it often misinterprets the intent of keyword-rich web-search queries, limiting the benefits of reasoning.

% keyword-rich, phrase-driven web-search queries
%struggles with keyword-style queries issued by agents and it  the reasoning advantages.
%web-search-style queries issued by agents make Rank1's reasoning less effective, and we think this is because Rank1 has not been trained on such query formats.
For RQ3, we propose a \textit{query-to-question} (Q2Q) method to translate agent-issued web search queries into natural-language questions (similar to MS MARCO-style questions~\citep{bajaj2018ms}), significantly improving neural retrieval and re-ranking performance. 
This indicates that the mismatch between agent-issued queries and queries used for training neural rankers can severely degrade neural ranking effectiveness.
Mitigating this training--inference query mismatch is therefore critical for improving neural rankers in deep research.

\header{Contributions}
Our main contributions are as follows:
\begin{itemize}[leftmargin=*,nosep]
\item To the best of our knowledge, we are the first to reproduce a comprehensive set of text ranking methods in deep research.

\item We construct a passage corpus for the recent deep research dataset BrowseComp-Plus.
\item Experiments across 2 open-source agents, 5 retrievers, and 3 re-rankers reveal the effectiveness of the following components in deep research: passage-level information units, retrievers suited to web-search-style queries, re-ranking, and mitigation of the training--inference query mismatch in neural ranking.
\item We open-source our code, data, and all agent-generated traces at \url{https://github.com/ChuanMeng/text-ranking-in-deep-research}.
%
%The traces are released in an encrypted format and can be locally decrypted for post-hoc analyses of agent behaviour.
% Following the BrowseComp-Plus protocol, 
%These traces can be used to support future post-hoc analyses of agent behaviour.
%\todo{They're sensitive about releasing the queries and such in an easy-to-use format (e.g., the decryption stuff), since they could be used to train LLMs. Are we sure it's OK to release this?}
%
\end{itemize}

\vspace*{-2mm}
\section{Task definition}
\vspace*{-0.3mm}

%\header{Traditional text ranking}
%Text ranking in ad-hoc search follows a single-shot paradigm over user-issued queries. 
%Given a query $q_u$ and corpus $C=\{d_1,\dots,d_n\}$, a ranking function $f$ returns a ranked list $D \subseteq C$ of size $k$, i.e., $D=f(q_u, C)$.

Text ranking has been extensively studied in ad-hoc search, which typically follows a single-shot paradigm over user-issued queries. 
Given a query $q_u$ issued by a user and a corpus of documents 
$C=\{d_1,d_2,\dots,d_n\}$ with $n$ documents, the goal of a text ranking method $f$ is to return a ranked list $D \subseteq C$ of size $k$, i.e., $D=f(q_u, C)$.

%\begin{equation}
%D = f(q_u).
%\label{eq:trad}
%\end{equation}

\header{Text ranking in deep research}
We follow the ReAct~\citep{yao2022react} paradigm, widely used in recent work~\citep{xu2026sage,chen2025browsecomp}, 
to define text ranking in deep research.
Given a query $q_u$ issued by a user, a deep research agent $A$ takes $q_u$ as input and performs 
multiple iterations of reasoning and search before producing the final answer $a$. 
At iteration $t$, the agent generates a reasoning trace $s_t$ that determines whether to output the final answer $a$ or invoke search to gather information. 
If the agent decides to invoke search, it issues a search query $q_t$ to a text ranking method $f$, 
which returns a ranked list $D_t$:
\begin{equation}
\begin{split}
\textstyle
q_t &= A(q_u, s_0, q_0, D_0, \ldots, s_t),\\
D_t &= f(q_t),
\end{split}
\label{eq:deep_rank}
\end{equation}
where $s_0$, $q_0$ are the reasoning trace and query generated by the agent at the initial iteration, respectively; $D_0$ is the ranked list returned by $f$ in response to $q_0$.
The ranked list $D_t$ with $k$ documents is then fed back to the agent to produce the reasoning trace 
at the next iteration $t+1$:
\begin{equation}
\textstyle
s_{t+1} = A(q_u, s_0, q_0, D_0, \ldots, s_t, q_t, D_t).
\label{eq:deep_state}
\end{equation}

\noindent Note that some agents may perform multiple consecutive search invocations or reasoning steps; for simplicity, the above definition assumes alternating reasoning and search steps.

\vspace*{-1.5mm}
\section{Methodology}
%\vspace*{-1.5mm}
\label{model}
%We state our research questions, the experiments designed to address them, and our experimental setup.
This section presents our research questions, experimental design, and experimental setup, including the agents and ranking methods, dataset, evaluation protocol, and our proposed method.

\vspace*{-1.5mm}
\subsection{Research questions and experimental design}
%\vspace*{-1.5mm}
% work is steered by the following research questions:
%\vspace*{-1.5mm}
\begin{enumerate}[label=\textbf{RQ\arabic*},leftmargin=*]
    \setItemnumber{1}
    \item To what extent are existing retrievers effective in deep research under passage-level and document-level retrieval units? \label{RQ1}
\end{enumerate}
\vspace*{-3mm}
To address \ref{RQ1}, we reproduce widely-used retrievers from different categories (see Section~\ref{sec_ranker}) in deep research, and compare their performance on passage and document corpora.
BrowseComp-Plus~\citep{chen2025browsecomp}, the dataset used in this study, provides only a document corpus; we construct a passage corpus (see Section~\ref{sec_passage_construct}). 
When using document-level retrieval units, prior work~\citep{chen2025browsecomp,sharifymoghaddam2026rerank} typically feeds truncated retrieved documents to agents to avoid exhausting the context window; \citet{chen2025browsecomp} further introduces a full-document reader tool that allows agents to access complete documents when needed.
Accordingly, we evaluate three settings: 
\begin{enumerate*}[label=(\roman*)]  
\item agents retrieve documents but read truncated versions;
\item same as (i), but with a full-document reader tool; and
\item agents retrieve and read passages.
In addition, we consider a setting in which agents retrieve and read passages, and can choose to invoke the full-document reader to read the source document of a retrieved passage.
\end{enumerate*}

\label{E1}
\begin{enumerate}[label=\textbf{RQ\arabic*},leftmargin=*]
    \setItemnumber{2}
    \item To what extent is a re-ranking stage effective in deep research under different initial retrievers, re-ranker types, and re-ranking cut-offs? \label{RQ2}
\end{enumerate}
\vspace*{-1mm}
To address \ref{RQ2}, we reproduce widely-used re-rankers~(see Section~\ref{sec_ranker}), and evaluate text ranking pipelines under various configurations, including different initial retrievers, different re-rankers, and different re-ranking depths.

\begin{enumerate}[label=\textbf{RQ\arabic*},leftmargin=*]
    \setItemnumber{3}
    \item To what extent does the mismatch between agent-issued queries and the training queries used for text ranking methods affect their performance?  \label{RQ3}
\end{enumerate}
\vspace*{-1mm} To address \ref{RQ3}, we compare the performance of text ranking methods using agent-issued search queries with their performance using natural-language questions similar to those in MS MARCO~\citep{bajaj2018ms}, on which many neural rankers are trained. 
We propose a query-to-question (Q2Q) method, which translates agent-issued web search queries into natural-language questions; see Section~\ref{sec_q2q} for details.

%based on an IR evaluation metric~(e.g., MAP@100, MRR@10). \label{E3}

\vspace*{-2mm}
\subsection{Experimental setup}

\subsubsection{Deep research agents}
We use two LLMs as deep research agents with model sizes that are feasible to a broad range of research groups.
Specifically, we follow~\citep{chen2025browsecomp} to use gpt-oss-20b~\citep{agarwal2025gpt}, and use the recently released GLM-4.7-Flash~\citep{zeng2025glm}.
Both LLMs have been trained to invoke web search, which is crucial for deep research.
%As shown in~\citep{chen2025browsecomp}, LLMs that have not been trained for tool use (e.g., Qwen3~\citep{yang2025qwen3}) generally rarely invoke search calls when explicitly prompted, leading to limited performance in deep research scenarios; we therefore do not consider such LLMs in this work.
\begin{itemize}[leftmargin=*,nosep]
\item \textbf{gpt-oss-20b}~\citep{agarwal2025gpt} is an OpenAI’s open-weight LLM.
It is pre-trained and subsequently post-trained for reasoning (i.e., using chain-of-thought) and tool use. For tool use, the model is trained to interact with the web via search and web page opening actions. It supports a maximum context window and output length of 131{,}072 tokens.
\item \textbf{GLM-4.7-Flash (30B)}~\citep{zeng2025glm} is an open-source LLM developed by Z.ai. 
It has a pre-training stage, followed by a mid-training phase (improves coding \& reasoning) and a post-training phase.
During post-training, the model is explicitly trained for web search via reinforcement learning (RL).
It supports a context window of 202{,}752 tokens and a maximum output length of 128{,}000 tokens.

\end{itemize}
Note that our work focuses on reproducing text ranking methods in the deep research scenario; considering other LLMs as deep research agents or scaling sizes is beyond the scope of this paper.

\subsubsection{Text ranking methods to be reproduced}
\label{sec_ranker}
We employ a set of widely-used and representative text-ranking methods in the IR community, including retrievers and re-rankers.

\header{Retrievers}
We use widely-used retrievers spanning 4 main paradigms
in modern IR: lexical-based sparse, learned sparse, single-vector dense, and multi-vector dense retrievers:
\begin{itemize}[leftmargin=*,nosep]
\item \textbf{BM25}~\citep{Robertson1994OkapiAT} is a lexical retriever based on vocabulary-level vectors using the bag-of-words approach for queries and documents.
\item \textbf{SPLADE-v3}~\citep{lassance2024splade} is a learned sparse retriever that trains BERT~\citep{devlin2019bert} to predict sparse vectors over BERT's vocabulary list for queries and documents.
\item \textbf{RepLLaMA}~\citep{ma2024fine} is a single-vector dense retriever; it fine-tunes Llama~2~\cite{touvron2023llama2} using LoRA~\cite{hu2021lora} to produce a single embedding vector for each query and document.
It appends an end-of-sequence token to each query or document input and uses the hidden state of the last model layer corresponding to this token as the embedding.
Its embedding dimension is 4,096.
\item \textbf{Qwen3-Embed-8B}~\citep{zhang2025qwen3} is a single-vector dense retriever. 
It belongs to the Qwen3-Embedding series (0.6B, 4B, and 8B), and we use the largest variant.
The series features fine-tuning of Qwen3 LLMs~\citep{yang2025qwen3} on synthetic training data across multiple domains and languages, generated by the Qwen3 models themselves.
It follows RepLLaMA to obtain query and document embeddings and shares the same embedding dimension.
%
%Its embedding dimension is 4,096.
%
\item \textbf{ColBERTv2}~\citep{santhanam2022colbertv2} is a multi-vector retriever. It trains BERT~\citep{devlin2019bert} to produce embeddings for each token in the query and the document, and models relevance as the sum of the maximum similarities between each query vector and all document vectors.

\end{itemize}
Note that the training data and input length configuration of Qwen3-Embed-8B differ substantially from those of other neural retrievers (SPLADE-v3, ColBERTv2, and RepLLaMA):
\begin{enumerate*}[label=(\roman*)]
\item Qwen3-Embed-8B is trained on fully synthetic data generated by Qwen3, whereas the others share the same training dataset, namely the training data of the MS MARCO V1 passage ranking corpus~\cite{bajaj2018ms};
\item Qwen3-Embed-8B is trained to support document lengths of up to 32,000 tokens, whereas the other neural ones are trained for passage-level inputs.
\end{enumerate*}
%
%We use 2 types of widely used pointwise re-rankers: non-reasoning-based and reasoning-based models.

\header{Re-rankers}
We select methods representing three effectiveness--efficiency trade-offs: a relatively inexpensive model (monoT5-3B~\citep{nogueira2020document}), an LLM-based re-ranker (RankLLaMA-7B~\citep{ma2024fine}), and a CoT-based reasoning re-ranker (Rank1-7B~\citep{weller2025rank}), which requires reasoning token generation.
All of them are pointwise re-rankers, which independently assign a relevance score given a query and a document:
\begin{itemize}[leftmargin=*,nosep]
\item \textbf{monoT5-3B}~\cite{nogueira2020document} is a non-reasoning-based re-ranker that fine-tunes T5~\cite{raffel2020exploring} to output either ``true'' or ``false'' to indicate relevance. 
The probability assigned to the ``true'' token is used as the relevance score. 
monoT5 is available in base (220M), large (770M), and 3B variants; we use the 3B model.
\item \textbf{RankLLaMA-7B}~\cite{ma2024fine} is a non-reasoning-based re-ranker that fine-tunes Llama~2~\cite{touvron2023llama2} with LoRA~\cite{hu2021lora} to project the representation of the end-of-sequence token to a relevance score.
\item \textbf{Rank1-7B}~\cite{weller2025rank} is a reasoning-based re-ranker that fine-tunes Qwen~2.5~\citep{yang2024qwen2} to generate a reasoning trace before producing a ``true''/``false'' decision (``<think> ... </think> true/false''). 
The ground truth training reasoning traces are generated by DeepSeek-R1~\citep{guo2025deepseek} on MS MARCO~\citep{bajaj2018ms}. 
As in monoT5, the probability assigned to the ``true'' token is used as the relevance score.
\end{itemize} 
All re-rankers are trained on the MS MARCO V1 passage ranking dataset~\cite{bajaj2018ms} and trained to operate on passage-level inputs. 
%
%We adopt pointwise ones because they offer competitive effectiveness while remaining computationally efficient~\citep{zhuang2024setwise}, making them more practical for deployment than their pairwise~\citep{qin2024large} and listwise~\citep{liu2025reasonrank,yang2025rank} counterparts.
Note that evaluating larger re-ranker variants (e.g., RankLLaMA-13B and Rank1-14B/32B) is beyond the scope of this work.

%RankLLaMA has 7B and 13B versions, and choose the 7B version because recent work~\citep{weller2025rank,yang2025rank} has shown that 13B has no obvious beneift than 7B and 7B peforms better than 13B sometimes.
%monoT5 and RankLLaMA are widely-used in the literature~\cite{weller2025rank,yang2025rank,meng2024ranked}, while Rank1~\cite{weller2025rank} is a recently proposed state-of-the-art reasoning-based re-ranker.
%%Rank1 has 7B, 14B, and 32B versions, a
%\item list one (optional)

\begin{table}[t!]
\centering
%\small
\caption{
Statistics of the BrowseComp-Plus dataset~\citep{chen2025browsecomp}.
Length is measured in tokens using the Qwen3~\citep{yang2025qwen3} tokeniser.
}
\label{tab_data}
\begin{tabular}{c c l c c}
\toprule
\# Q & Avg.\ Q Len. & Corpus Type & \# Items & Avg.\ Item Len. \\
\midrule
\multirow{2}{*}{830} & \multirow{2}{*}{132.19} & Document & \phantom{0,}100{,}195    & 7{,}845.55 \\
                     &                        & Passage  & 2{,}772{,}255 & \phantom{0,}279.64     \\
\bottomrule
\end{tabular}
%\vspace*{-2mm}
\end{table}

\subsubsection{Dataset.}
We use BrowseComp-Plus~\citep{chen2025browsecomp}, a deep research dataset built on BrowseComp~\citep{wei2025browsecomp}, which comprises 1,266 fact-seeking, reasoning-intensive, long-form queries together with their answers; the answers are generally short and objective, making answer evaluation easier.
%; queries and answers are self-contained.
%
BrowseComp-Plus adds a document corpus and human-verified relevance judgments via a three-stage process.
\begin{enumerate*}[label=(\roman*)]
\item each question--answer pair is sent to OpenAI o3 to identify clues (a clue is a part of the question) useful for deriving the answer and to return supporting web documents;
\item human annotators verify whether each clue is well supported by its supporting documents and whether the combination of clues and documents enables answering the question; annotators revise the clues and supporting documents when necessary;
\item the final document corpus is constructed by including all human-verified supporting documents, together with mined hard-negative documents.
\end{enumerate*}
Some queries are removed in (i) and (ii) when OpenAI o3 fails to return valid clues/supporting documents, or it is too hard for human annotators to correct them, resulting in a final set of 830 queries.
The dataset provides two types of relevance judgments: \textit{gold} and \textit{evidence}.
Gold documents contain the final answers (not necessarily exact matches) to the queries, while evidence documents include the gold documents and documents supporting intermediate reasoning steps.
On average, each query has 2.9 gold, 6.1 evidence, and 76.28 negative documents.
Table~\ref{tab_data} reports detailed statistics.

\subsubsection{Passage corpus construction}
\label{sec_passage_construct}
To segment the documents into passages, we follow~\citep{Owoicho2022TRECC2,Dalton2021TRECC2} to split each document in the original document corpus of BrowseComp-Plus into canonical passages of at most 250 words using the spaCy toolkit with the ``en\_core\_web\_sm'' model; we use the publicly available code.\footnote{\url{https://github.com/grill-lab/trec-cast-tools/tree/master/corpus_processing}}
The statistics of the passage corpus are shown in Table~\ref{tab_data}.
Each passage is assigned a new passage ID, and we record the mapping from each passage to its original document.
Following~\citep{dai2019deeper}, when a document title is available, we extract it and prepend it to the beginning of each corresponding passage to provide additional context.

\subsubsection{Query-to-question (Q2Q) reformulation}
\label{sec_q2q}
To translate a web search query $q_t$ issued by an agent at iteration $t$ into a natural-language question, we define a query-to-question (Q2Q) reformulator $g$ that takes $q_t$ as input and outputs a natural-language question $\tilde{q}_t$, i.e., $\tilde{q}_t = g(q_t)$; then, $\tilde{q}_t$ is sent to a text ranking method $f$ to return a ranked list $D_t = f(\tilde{q}_t)$.
However, a web search query $q_t$ may be ambiguous and may not clearly reflect the agent’s search intent; relying solely on the query may therefore cause the reformulation to deviate from the agent’s search intent.
To provide additional context about the agent’s search intent, we introduce another variant of Q2Q that includes the recent reasoning trace $s_t$ generated by the agent, namely $\tilde{q}_t = g(q_t, s_t)$.

%Document titles can be prepended to their corresponding passages to provide additional context.

%We follow~\citep{dai2019deeper}, when the title is available, to prepend the document title to the beginning of every passage belonging to that document to provide additional context.
%

%en_core_web_-
%sm-3.0.0 model. 
%and the en_core_web_sm-3.3.0 model

\subsubsection{Evaluation.}
We follow the original BrowseComp-Plus evaluation protocol~\citep{chen2025browsecomp} and report \textit{the number of search calls}, \textit{recall}, and \textit{accuracy}; we use the evaluation code released by the dataset authors.\footnote{\url{https://github.com/texttron/BrowseComp-Plus}}
Search calls denote the average number of search invocations per query.
Recall measures the proportion of evidence documents returned across all search calls for a query.
Accuracy is computed using an LLM-as-judge that compares an agent’s final answer with the ground-truth answer.
To analyse token-budget efficiency, we additionally report \textit{completion rate}, i.e., the percentage of queries for which the agent outputs a final answer before reaching the maximum context-window or output-token limits; queries reaching these limits receive accuracy 0.

\header{Evaluating on passage corpus}
Because BrowseComp-Plus provides relevance judgments at the document level, they cannot be directly applied to the passage corpus.
When evaluating passage retrieval, we therefore adopt the \textit{Max-P} strategy~\citep{dai2019deeper}, mapping retrieved passages to documents by assigning each document the maximum score amongst its retrieved passages.

\begin{table}[t!]
\centering
%\small
\caption{
Sanity-check comparison of agent performance for gpt-oss-20b using Qwen3-Embed-8B as a search tool on the original BrowseComp-Plus document corpus. 
Results from two recent studies~\citep{chen2025browsecomp,sharifymoghaddam2026rerank} are shown alongside our replicated results.
We also report results for GPT-5.2 (high reasoning mode), which tends to generate longer reasoning traces and search much more aggressively, causing 354 of 830 queries to reach the maximum iteration limit (100); these queries are assigned an accuracy of 0, resulting in lower overall accuracy.
}
\label{tab_replicate}
\begin{tabular}{l llll}
\toprule
Agent & Search calls & Recall & Acc. & vLLM\\
\midrule
gpt-oss-20b~\citep{chen2025browsecomp} & 23.87  & 0.493 & 0.346 & v0.9.0.1\\
gpt-oss-20b ~\citep{sharifymoghaddam2026rerank} & 29.63 & 0.557 & 0.422 & v0.13\\
\hline
gpt-oss-20b (ours) & 30.14 & 0.570 & 0.421 &v0.15 \\
GPT-5.2 (ours)  & 73.83 & 0.747 & 0.451  & \phantom{00}- \\
\bottomrule
\end{tabular}
\end{table}

\subsubsection{Implementation details.}
\label{sec:details}

Regarding the agent setup, for both gpt-oss-20b\footnote{\url{https://huggingface.co/openai/gpt-oss-20b}} and GLM-4.7-Flash (30B)\footnote{\url{https://huggingface.co/zai-org/GLM-4.7-Flash}}, following~\citep{chen2025browsecomp}, we set the maximum output length to 40{,}000 tokens and the maximum number of iterations to 100. 
We run gpt-oss-20b in the high reasoning-effort mode (the default setting); the reasoning effort of GLM-4.7-Flash (30B) is not configurable.
Both agents are deployed locally using vLLM version 0.15, which satisfies the minimum version requirement of GLM-4.7-Flash.

Regarding the text ranking setup, we follow prior work~\citep{sharifymoghaddam2026rerank,chen2025browsecomp}: after each search call, we return the top-5 ranked documents (or passages) to the agent and truncate each returned document to its first 512 tokens; this prevents long documents from exhausting the context window: with an average document length of 8K tokens (Table~\ref{tab_data}), a 130K-token context window (gpt-oss-20b) could hold only about 16 full documents.
We use Pyserini's BM25 with its default parameters ($k_1=0.9$, $b=0.4$), following~\citep{chen2025browsecomp}.
We implement SPLADE-v3 (naver/splade-v3), RepLLaMA (castorini/repllama-v1-7b-lora-passage)\footnote{We also evaluated the RepLLaMA checkpoint trained on the MS MARCO document corpus (castorini/repllama-v1-7b-lora-doc); however, it performed worse on BrowseComp-Plus than the passage-trained checkpoint (repllama-v1-7b-lora-passage).}, RankLLaMA-7B (castorini/rankllama-v1-7b-lora-passage), and Qwen3-Embed-8B (Qwen/Qwen3-Embedding-8B) using Tevatron.\footnote{\url{https://github.com/texttron/tevatron}}
We implement ColBERTv2 (colbert-ir/colbertv2.0) using PyLate~\citep{chaffin2025pylate}.\footnote{\url{https://github.com/lightonai/pylate}}
We implement monoT5-3B (monot5-3b-msmarco) following PyTerrier\_t5.\footnote{\url{https://github.com/terrierteam/pyterrier_t5}}
Rank1-7B (jhu-clsp/rank1-7b) is implemented following the original authors’ implementation.\footnote{\url{https://github.com/orionw/rank1}}
For document indexing, we follow~\citep{chen2025browsecomp} to set the maximum input length of Qwen3-Embed-8B to 4,096 tokens; the remaining neural retrievers use a maximum length of 512 tokens, as they are trained on passage-level inputs.
For passage indexing, all retrievers use a maximum length of 512 tokens.
We implement the query-to-question (Q2Q) reformulator (see Section~\ref{sec_q2q}) using gpt-oss-20b in the low reasoning-effort mode; we randomly sample natural language questions from TREC-DL 2019~\citep{craswell2019} and include them in the prompt as examples to specify the desired question-style output.
Experiments are conducted using NVIDIA RTX 6000 Ada (48 GB), H100 (80 GB), and H200 (141 GB) GPUs, subject to hardware availability. 

%\citep{chen2025browsecomp} show that when documents are truncated to the first 512 tokens, 86.5\% of queries still contain the ground-truth answer in at least one of their gold documents).

%gpt-oss-20b-high
%gpt-oss-120B-high
%gpt-5.2
%Seed-OSS-36B-Instruct~\footnote{\url{https://huggingface.co/ByteDance-Seed/Seed-OSS-36B-Instruct}}

%input window: 128K
%output limit: 40K

%512 (one doc)*5(number of doc)*50 (iteractions)=128000

%key not k
\begin{table*}[ht!]
\centering
\caption{
Agent performance across retrievers on BrowseComp-Plus.
The agent is based on gpt-oss-20b; it supports a maximum context window of 131{,}072 tokens.
``\#Search'' and ``\#GetDoc'' denote the number of search and full-document reader calls, respectively.
``Compl.'' denotes the completion rate, i.e., the percentage of queries for which the agent outputs a final answer before reaching the maximum context-window or output-token limits; queries reaching these limits receive accuracy 0.
The best values in recall and accuracy are \textbf{boldfaced}, and the second-best ones are \underline{underlined}.
}
\label{tab_gpt_oss_retrieval}
\setlength{\tabcolsep}{4.5pt} 
\begin{tabular}{l cccc cccc ccccc}
\toprule
\multirow{2}{*}{Retriever} 
& \multicolumn{4}{c}{Passage corpus} 
& \multicolumn{4}{c}{Document corpus}  
& \multicolumn{5}{c}{Document corpus+GetDoc} \\
\cmidrule(lr){2-5} \cmidrule(lr){6-9} \cmidrule(lr){10-14}
& \#Search & Recall & Acc. & Comp.
& \#Search  & Recall & Acc. & Comp.
& \#Search  & \#GetDoc & Recall & Acc. & Comp. \\
\midrule
BM25 & 30.97 & \textbf{0.616}  & \textbf{0.572}  & 0.968 & 32.22  & 0.366   & 0.259 &  0.805  & 28.14  & 1.31 & 0.343 & 0.301 & 0.746 \\
SPLADE-v3  & 32.75  & 0.545  & 0.516 & 0.980   & 28.95  &  \underline{0.628}  & \underline{0.476}  & 0.824 & 24.64 & 1.65 & \underline{0.602} & \underline{0.529} & 0.829 \\
RepLLaMA & 36.13 & 0.449  & 0.406 & 0.961 & 30.69 & 0.514 & 0.363 & 0.786 & 26.86 & 1.46 & 0.476 & 0.399 & 0.816 \\
Qwen3-Embed-8B  & 37.83 & 0.470 & 0.417 & 0.963  & 30.14  & 0.570 & 0.421  & 0.821  & 26.56 & 1.63 & 0.559 & 0.455 & 0.823 \\
ColBERTv2 & 32.88 & \underline{0.552}  & \underline{0.521} & 0.978 & 27.13 & \textbf{0.633} & \textbf{0.481} & 0.855 & 24.31 & 1.75 & \textbf{0.612} & \textbf{0.538} & 0.835 \\
\bottomrule
\end{tabular}
\end{table*}

\begin{table*}[t!]
\centering
\caption{
Agent performance across different retrievers on BrowseComp-Plus.
The agent is based on GLM-4.7-Flash (30B); it supports a maximum context window of 200{,}000 tokens.
%
%``\#Search'' denotes the number of search calls.
%
%``Compl.'' denotes the completion rate.
%(see caption in Table~\ref{tab_gpt_oss_retrieval} for details).
%
%The most effective values in recall and accuracy are \textbf{boldfaced}, and the second-best ones are \underline{underlined}.
Metric definitions and formatting follow Table~\ref{tab_gpt_oss_retrieval}.
}
\label{tab_glm_retrieval}
\setlength{\tabcolsep}{4pt} 
\begin{tabular}{l cccc cccc ccccc}
\toprule
\multirow{2}{*}{Retriever} 
& \multicolumn{4}{c}{Passage corpus} 
& \multicolumn{4}{c}{Document corpus}  
& \multicolumn{5}{c}{Document corpus+GetDoc} \\
\cmidrule(lr){2-5} \cmidrule(lr){6-9} \cmidrule(lr){10-14}
& \#Search  & Recall & Acc. & Comp.
& \#Search & Recall & Acc. & Comp.
& \#Search & \#GetDoc & Recall & Acc. & Comp. \\
\midrule
BM25 & 34.73 & \textbf{0.581} & 0.445 & 0.863 & 26.83 & 0.309 & 0.196 & 0.923 & 21.86 & 3.00  & 0.282 & 0.263  & 0.947 \\
%ColBERTv2 &  &  & & &  &  & &  &  &  &  &  & \\
SPLADE-v3  & 36.36 & \underline{0.578} & \textbf{0.466} & 0.883 & 25.62 & \underline{0.639} & \textbf{0.448} & 0.881 & 18.43  & 5.33 & \textbf{0.597} & \underline{0.525} & 0.939\\
RepLLaMA & 37.60 & 0.456 & 0.331 & 0.905 & 25.77 & 0.493 & 0.330 & 0.907 & 20.93 & 4.13 & 0.471  & 0.407  & 0.955 \\
Qwen3-Embed-8B  & 37.47 & 0.482 & 0.357 & 0.886 & 26.25 & 0.580 & 0.374 & 0.882 & 20.12 & 4.34  & 0.482  & 0.456 & 0.956 \\
ColBERTv2  & 36.41 & 0.571 & \underline{0.464}  & 0.882 &  26.06 & \textbf{0.640} & \underline{0.430} & 0.878 & 18.85 &  5.30 &  \underline{0.595} & \textbf{0.535}  & 0.955 \\
\bottomrule
\end{tabular}
\end{table*}

\begin{table}[h!]
\centering
\caption{
Examples of three search queries issued by the gpt-oss-20b and GLM-4.7-Flash agents for Query 781 on BrowseComp-Plus.
}
\label{table_query_example}
%\small
\setlength{\tabcolsep}{1.2pt} 
\begin{tabular}{l l}
\toprule
Agent & Search query \\
\midrule
gpt-oss-20b  & ``90+7'' attendance 61700 \\
            & ``Man United'' ``4-1'' ``90+4''\\
            & ``assist'' ``4--1'' ``Premier League'' ``2020'' \\
\midrule
GLM-4.7-Flash & ``90'+4'' football match attendance \\
                    & ``61,888'' football match Stockholm Vienna Prague \\
                    &  ``goal in the 6th minute'' football\\
\bottomrule
\end{tabular}
\end{table}

\begin{table}[h!]
\centering
\caption{
Performance of the gpt-oss-20b agent across different retrievers on BrowseComp-Plus.
%
%``\#Search'' denotes the number of search calls.
%
%``Compl.'' denotes the completion rate.
%
%The most effective values in recall and accuracy are \textbf{boldfaced}, and the second-best ones are \underline{underlined}.
Metric definitions and formatting follow Table~\ref{tab_gpt_oss_retrieval}.
}
\label{tab_gpt_oss_retrieval_passage_getdoc}
\setlength{\tabcolsep}{4pt} 
\begin{tabular}{l ccccc }
\toprule
\multirow{2}{*}{Retriever} 
& \multicolumn{4}{c}{Passage corpus+GetDoc} \\
\cmidrule(lr){2-6} 
& \#Search  & \#GetDoc   & Recall & Acc. & Comp. \\
\midrule
BM25 & 26.87 & 2.24  & \textbf{0.556}  & \textbf{0.542} &  0.877 \\
SPLADE-v3   & 27.54  & 2.11  & \underline{0.532}   & \underline{0.510} & 0.895 \\
RepLLaMA  & 31.82 & 1.77  & 0.399   & 0.369 & 0.868\\
Qwen3-Embed-8B  & 32.11  &  1.98  & 0.438  &  0.404 & 0.853\\
\bottomrule
\end{tabular}
\end{table}

\vspace*{-1.5mm}
\section{Results and Discussions}

\vspace*{-1mm}
\subsection{Sanity check for reproduction}
\label{sec_replicate}
As a sanity check, we attempt to replicate the results reported in recent studies~\citep{chen2025browsecomp,sharifymoghaddam2026rerank}.
We evaluate gpt-oss-20b (high-reasoning mode) with Qwen3-Embed-8B as the retriever on the original BrowseComp-Plus document corpus. The results are shown in Table~\ref{tab_replicate}.
Our results obtained with the latest vLLM version (v0.15) are very similar to those produced using v0.13 in \citep{sharifymoghaddam2026rerank}.
We observe that newer vLLM versions (v0.15 and v0.13) lead to higher performance compared to the old one (v0.9.0.1).
Besides replicating gpt-oss-20b, we also test GPT-5.2, a current state-of-the-art commercial LLM.
We find that GPT-5.2 requires substantially more iterations; 354 queries (830 queries in total) require over 100 iterations. 
Such a large number of iterations would incur significant API costs, making it less feasible for broader researchers and practitioners (all queries cost roughly \$1,000 to \$2,000); therefore, we do not use it in this work.

\vspace*{-1.5mm}
\subsection{Retrievers on passage and document corpora}
\label{sec_rq1}
%\vspace*{-1.5mm}

To answer \ref{RQ1}, we compare two agents, gpt-oss-20b and GLM-4.7-Flash (30B), across different retrievers on both the passage and document corpora.
For experiments on the document corpus, following~\citep{chen2025browsecomp,sharifymoghaddam2026rerank}, we truncate each retrieved document to the first 512 tokens, to avoid exhausting the input context window.
To mitigate truncation-induced information loss, we follow~\citep{chen2025browsecomp} to evaluate a setting where agents can access a full-document reader tool on the document corpus.
We present the results in Tables~\ref{tab_gpt_oss_retrieval} (gpt-oss-20b) and~\ref{tab_glm_retrieval} (GLM-4.7-Flash).
We have four main observations.

%although agents on the document corpus generally achieve higher retrieval performance (recall).
First, both agents achieve higher answer accuracy on the passage corpus than on the document corpus (without the full-document reader) across all retrievers, except Qwen3-Embed-8B, likely because it is trained on substantially longer documents (up to 32K tokens) and is less effective on passages due to distribution shift.
Notably, the relative improvement on passages is more pronounced for the shorter-context gpt-oss-20b (131K context window) than for GLM-4.7-Flash (200K).
E.g., with SPLADE-v3, gpt-oss-20b achieves 0.516 accuracy on passages, an 8.4\% relative improvement over documents (0.476), whereas GLM-4.7-Flash gains 4.02\%.
We further observe that both agents issue more search calls on passages than on documents (e.g., 32.75 vs. 28.95 with gpt-oss-20b using SPLADE-v3). 
Moreover, for gpt-oss-20b, the completion rate is markedly higher on passages (0.980 vs. 0.824), while GLM-4.7-Flash shows similar completion rates across corpora; note that both models share the same output-token limit but differ in context-window limit.
Taken together, these results suggest that passage-level units enable more search and reasoning iterations before reaching context-window limits, which in turn improves answer accuracy, particularly for agents with smaller context windows.

Second, when paired with the lexical retriever BM25, both agents achieve highly competitive performance on the passage corpus compared with neural rankers. 
For instance, gpt-oss-20b with BM25 attains the highest recall (0.616) and answer accuracy (0.572) on the passage corpus. 
The 0.572 accuracy is also the highest across all retrieval settings in Tables~\ref{tab_gpt_oss_retrieval} and \ref{tab_glm_retrieval}.
Table~\ref{table_query_example} presents examples of agent-issued queries; they follow a web-search style, characterised by keywords, phrases, and quotation marks for exact matching, making them well suited to lexical retrievers such as BM25.
In contrast, BM25 performs worst on the document corpus; we analyse this issue in more detail later in this section.

Third, single-vector dense retrievers (RepLLaMA, Qwen3-Embed-8B), despite their substantially larger model sizes, consistently underperform smaller BERT-based~\citep{devlin2019bert} learned-sparse (SPLADE-v3) and multi-vector dense (ColBERTv2) retrievers.
This finding is consistent with prior work showing that SPLADE and ColBERT perform well across diverse query formats on the BEIR dataset~\citep{thakur2021beir,formal2021spladev2} that include keyword-rich queries or those requiring exact matching.
ColBERT, in particular, has been shown to exhibit strong preferences for exact matching~\citep{mueller2025semantically}.
In contrast, single-vector approaches often struggle to adapt to new tasks and domains~\citep{thakur2021beir} and are theoretically constrained in representational capacity~\citep{weller2025theoretical}.

Fourth, enabling the full-document reader on the document corpus reduces search calls and recall but improves answer accuracy. 
E.g., with gpt-oss-20b using SPLADE-v3, accuracy increases from 0.476 to 0.529.
%, comparable to 0.516 on the passage corpus.
%
This suggests that the reader compensates for information loss caused by document truncation.
With the reader enabled on documents, gpt-oss-20b achieves performance comparable to the passage setting, while GLM-4.7-Flash generally surpasses the passage setting and maintains high completion rates, likely due to its longer context window, which allows processing of more complete documents.
We also evaluate enabling the reader on the passage corpus using gpt-oss-20b with several retrievers; see results in Table~\ref{tab_gpt_oss_retrieval_passage_getdoc}. 
We find the reader slightly degrades performance (e.g., with BM25, accuracy decreases from 0.572 to 0.542), likely because passage-level units already provide direct access to relevant segments within a document, rendering the reader redundant.

\begin{figure*}[ht!]
    \centering
    \begin{subfigure}{0.51\columnwidth}
        \includegraphics[width=\linewidth,trim=4mm 3mm 4mm 4mm,clip]{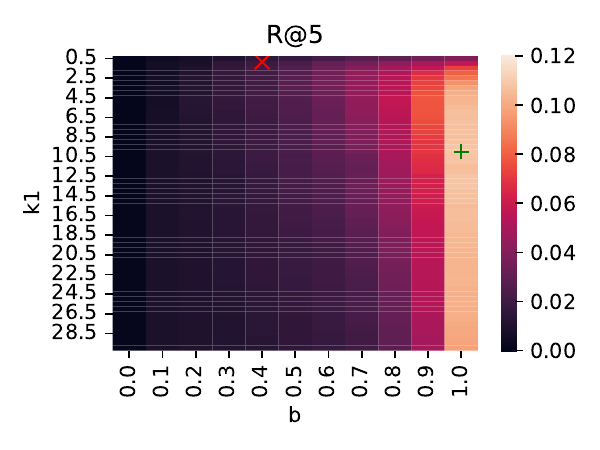}
        \vspace*{-6mm}
        \caption{Document corpus (Recall@5)}
        \label{fig_bm25_parameter_doc_r5}
    \end{subfigure}
    \begin{subfigure}{0.51\columnwidth}
        \includegraphics[width=\linewidth,trim=4mm 3mm 4.5mm 4mm,clip]{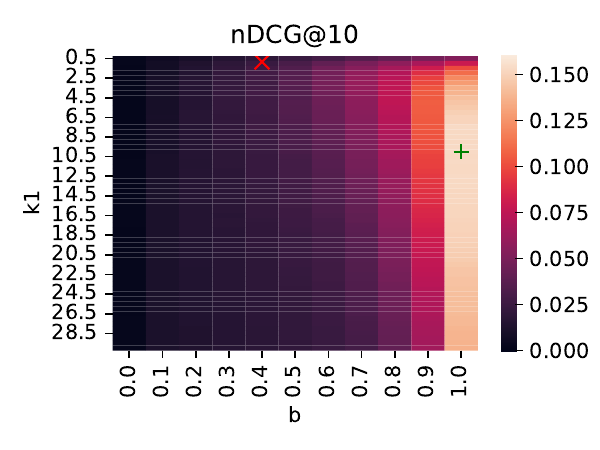}
        \vspace*{-6mm}
        \caption{Document corpus (nDCG@10)}
        \label{fig_bm25_parameter_doc_ndcg5}
    \end{subfigure}
    \begin{subfigure}{0.51\columnwidth}
        \includegraphics[width=\linewidth,trim=4mm 3mm 4.5mm 4mm,clip]{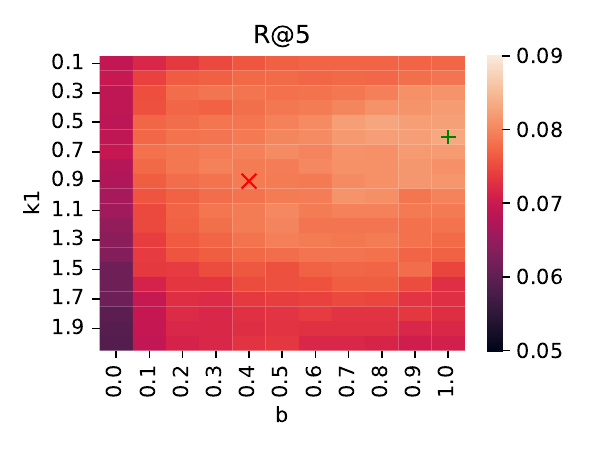}
        \vspace*{-6mm}
        \caption{Passage corpus (Recall@5)}
        \label{fig_bm25_parameter_psg_r5}
    \end{subfigure}
    \begin{subfigure}{0.51\columnwidth}
        \includegraphics[width=\linewidth,trim=4mm 3mm 4.5mm 4mm,clip]{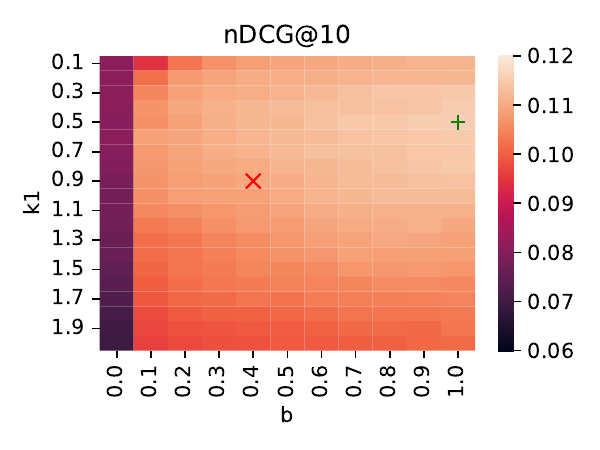}
        \vspace*{-6mm}
        \caption{Passage corpus (nDCG@10)}
        \label{fig_bm25_parameter_psg_ndcg5}
    \end{subfigure}
    \caption{
    Heatmap from a grid search on BrowseComp-Plus using the original full queries (not end-to-end), showing the effectiveness (evaluated by evidence judgments) of BM25 under different hyperparameter settings.
    The \textcolor{red}{red $\times$} denotes the default parameter setting following \citep{chen2025browsecomp}, while the \textcolor{ForestGreen}{green $+$} denotes the best parameter setting found by the grid search.
    The lighter the colour, the higher the retrieval performance.
    }
    \label{fig_bm25_parameter}
\end{figure*}

\vspace*{1mm}\noindent\textbf{Why does BM25 perform poorly on the document corpus?}
Prior work~\citep{kaszkiel1997passage} shows that lexical document retrieval is sensitive to document-length normalisation. 
BM25 has two parameters $k_1$ and $b$~\citep{Robertson1994OkapiAT}: $k_1$ controls term-frequency saturation (larger $k_1$ values lead to slower term frequency saturation, giving repeated terms more weight), while $b$ controls document length normalisation (larger $b$ values increase the penalty on longer documents).
To investigate this issue, we evaluate the gpt-oss-20b agent using BM25 under two settings:
\begin{enumerate*}[label=(\roman*)]
\item truncating each document to its first 512 tokens before indexing, thereby reducing the impact of document-length variation; and
\item instead of using the default parameter ($k_1=0.9$, $b=0.4$) following~\citep{chen2025browsecomp}, we use document-retrieval-oriented BM25 parameters ($k_1=3.8$, $b=0.87$), which increase length normalisation and have been shown to be effective for the MS MARCO document retrieval task\footnote{\url{https://github.com/castorini/anserini/blob/master/docs/experiments-msmarco-doc.md}}
\end{enumerate*}.
We present the results in Table~\ref{table_bm25_parameter}. 
Indexing only the first 512 tokens of each document substantially improves performance, yielding a 64.2\% relative gain in recall and a 98.1\% gain in answer accuracy, suggesting that reducing the effect of document-length normalisation benefits BM25 on documents. 
Using the document-oriented BM25 parameters also improves performance, with a 76.8\% gain in recall and a 71.0\% gain in accuracy.

To better understand the optimal BM25 parameter settings, we perform a grid search of retrieval performance on BrowseComp-Plus using the original full queries across different parameter values.
The results are shown in Figure~\ref{fig_bm25_parameter}.
We find that larger values of $b$ generally improve performance, i.e., penalising long documents with higher term frequencies is beneficial.
Moreover, the performance gap between the default setting ($k_1 = 0.9$, $b = 0.4$) and the optimal parameter settings is substantially larger on the document corpus than on the passage corpus. 
E.g., on the document corpus, $k_1 = 10$ and $b = 1$ appear to be a sweet spot (see Figures~\ref{fig_bm25_parameter_doc_r5} and \ref{fig_bm25_parameter_doc_ndcg5}), and performance using it is substantially different from the default setting.
We further evaluate gpt-oss-20b using BM25 with $k_1 = 10$ and $b = 1$.
Table~\ref{table_bm25_parameter} shows that this configuration achieves the highest recall (0.647) across all settings in this section.

We note that $k_1$ and $b$ are generally regarded as corpus-dependent parameters rather than query-specific ones~\cite{DBLP:conf/ecir/HeO05}. 
Indeed, Figure~\ref{fig_bm25_parameter} shows that strong length normalisation ($b=1$) and large $k_1$ values are important for effective BM25 retrieval over BrowseComp-Plus documents.
Precise tuning of these parameters is not required for strong effectiveness; in particular, a wide range of $k_1$ values performs well.
Nonetheless, we urge readers to exercise some caution when comparing the tuned BM25 results with other retrievers, since the parameter selection was performed directly on all BrowseComp-Plus queries (due to the absence of a validation set).

%We note that $k_1$ and $b$ are generally seen as settings that characterise properties about the collection itself, rather than topic-specific ones~\cite{DBLP:conf/ecir/HeO05}. 
%
%Indeed, Figure~\ref{fig_bm25_parameter} shows that length normalisation ($b=1$) and high term saturation ($k_1>5$) are important for effective BM25 retrieval over BrowseComp-Plus documents. 
%We note that it is not necessary to tune these parameters just right to get strong effectiveness: in particular, a wide range of $k_1$ values works well.
%Nonetheless, we urge readers to exercise some caution when comparing the tuned BM25 results with other retrievers, since the parameter selection was performed directly on all BrowseComp-Plus queries (due to the absence of a validation set).

\begin{table}[t!]
\centering
\caption{
Performance of the gpt-oss-20b agent using BM25 under different hyperparameter settings on BrowseComp-Plus.
Index len.\ indicates which portion of each document is indexed, i.e., the full document or the first 512 tokens.
$k_1 = 0.9$ and $b = 0.4$ are the default BM25 parameters, following~\citep{chen2025browsecomp}.
The best values in recall and accuracy are \textbf{boldfaced}, and the second-best ones are \underline{underlined}.
}
\label{table_bm25_parameter}
\setlength{\tabcolsep}{3pt} 
\begin{tabular}{l l cccc}
\toprule
\multirow{2}{*}{BM25 setting} 
& \multicolumn{4}{c}{Document corpus} \\
\cmidrule(lr){2-6}
&  Index len.  & \#Search  & Recall & Acc. & Comp. \\
\midrule
$k_1=0.9$, $b=0.4$ & full doc & 32.22 & 0.366 & 0.259 & 0.805 \\ 
$k_1=0.9$, $b=0.4$ &  512 token  &  28.00 & \underline{0.642}  & \textbf{0.513} & 0.812 \\
%$k_1=0.7$, $b=0.5$  & full doc  & 31.08 & 0.407  & 0.299 & 0.805 \\
%$k_1=4.46$, $b=0.82$ & full doc & 28.66 & 0.601  & 0.443 & 0.839 \\ 
$k_1=3.8$, $b=0.87$ & full doc & 28.66 & 0.601  & 0.443 & 0.839 \\ 
$k_1=10$, $b=1$ & full doc & 28.73 & \textbf{0.647} & \underline{0.506} & 0.848 \\
\bottomrule
\end{tabular}
\end{table}

\begin{table*}[th!]
\centering
\caption{
Performance of the gpt-oss-20b agent across ranking pipelines with different retrievers, re-rankers, and re-ranking depths on the passage corpus of BrowseComp-Plus.
$d$ denotes the re-ranking depth.
The best values in recall and accuracy are \textbf{boldfaced}, and the second-best ones are \underline{underlined}.
}
\label{tab_rerank_oss}
%\small
\begin{tabular}{l cccc cccc cccc}
\toprule
\multirow{2}{*}{Re-ranker} 
& \multicolumn{4}{c}{BM25} 
& \multicolumn{4}{c}{SPLADE-v3}  
& \multicolumn{4}{c}{Qwen3-Embed-8B} \\
\cmidrule(lr){2-5} \cmidrule(lr){6-9} \cmidrule(lr){10-13}
& \#Search  & Recall & Acc. & Comp.
& \#Search  & Recall & Acc. & Comp.
& \#Search  & Recall & Acc. & Comp. \\
\midrule
no re-ranking & 30.97 & 0.616 & 0.572  & 0.980 & 32.75 & 0.545 & 0.516  & 0.980   & 37.83  & 0.470 & 0.417 & 0.963 \\
\midrule
monoT5 ($d=10$) & 29.89 & 0.660 & 0.631 & 0.971 & 31.58 &  0.630 & 0.599 & 0.965  &  35.64 & 0.524 & 0.471 &  0.969 \\
monoT5 ($d=20$) & 28.96 & 0.701 & 0.674 & 0.960  & 30.07 & 0.647 & 0.598 &  0.974 & 34.16  & 0.570 & 0.541 & 0.959 \\
monoT5 ($d=50$) & 27.57  & \textbf{0.716}  & \textbf{0.689} & 0.974  &  29.72 & 0.689  & \underline{0.646} &  0.971  & 32.94 & \underline{0.614} & 0.559 & 0.952  \\
\hline
RankLLaMA ($d=10$) & 29.85 & 0.657 & 0.617 & 0.980 & 31.30 & 0.632 & 0.595 & 0.964 & 36.07 & 0.508 & 0.465 & 0.964 \\
RankLLaMA ($d=20$) & 29.03 & 0.681 & 0.655 & 0.971 & 30.82 & 0.646 & 0.605 & 0.981 & 34.25 & 0.553 & 0.494 & 0.961 \\
RankLLaMA ($d=50$) & 27.10 & 0.710 & 0.678 & 0.977 & 28.96 & \underline{0.691} & \textbf{0.663} & 0.959 & 32.85 & 0.613 & \textbf{0.568} & 0.964 \\
\hline
Rank1 ($d=10$) & 29.62 & 0.662 & 0.628 & 0.966 & 31.18 & 0.617 & 0.580 & 0.978 & 35.50 & 0.530 & 0.454 & 0.951 \\
Rank1 ($d=20$)& 28.28 & 0.694 & 0.669 & 0.971 & 30.73 & 0.672 & 0.617 & 0.953 & 34.04 & 0.579 & 0.528 & 0.969 \\
Rank1 ($d=50$)& 26.92 & \underline{0.712} & \underline{0.687} & 0.977 & 29.12 & \textbf{0.702} & 0.643 & 0.959 & 32.72 & \textbf{0.630} & \underline{0.564} & 0.949 \\
\bottomrule
\end{tabular}
\end{table*}

\begin{table}[th!]
\centering
\caption{
Performance of the GLM-4.7-Flash (30B) agent on the passage corpus of BrowseComp-Plus.
Metrics and formatting follow Table~\ref{tab_rerank_oss}.
%The most effective values in recall and accuracy are \textbf{boldfaced}, and the second-best ones are \underline{underlined}.
}
\label{tab_rerank_glm}

% \footnotesize
\setlength{\tabcolsep}{4pt}

\begin{tabular}{p{1cm} p{2.7cm} c c c c}
\toprule
Retriever & Re-ranker & \#Search & Recall & Acc. & Comp. \\
\midrule

\multirow{5}{*}{BM25}
& no re-ranking       & 34.73 & 0.581 & 0.445 & 0.863 \\
& monoT5 ($d=10$)      & 33.41 & 0.665 & 0.549 & 0.892 \\
& monoT5 ($d=50$)      & 31.46 & \underline{0.696} & \textbf{0.586} & 0.901 \\
& RankLLaMA ($d=10$)   & 37.99 & 0.660 & 0.503 & 0.836 \\
& RankLLaMA ($d=50$)   & 32.97 & \textbf{0.707} & \underline{0.576} & 0.865 \\
\midrule

\multirow{5}{*}{\shortstack[l]{SPLADE\\-v3}}
& no re-ranking       & 36.36 & 0.578 & 0.466 & 0.883 \\
& monoT5 ($d=10$)      & 32.24 & 0.632 & 0.528 & 0.865 \\
& monoT5 ($d=50$)      & 31.05 & \underline{0.693} & \textbf{0.575} & 0.900 \\
& RankLLaMA ($d=10$)   & 33.97 & 0.612 & 0.490 & 0.880 \\
& RankLLaMA ($d=50$)   & 34.62 & \textbf{0.695} & \underline{0.543} & 0.842 \\
\bottomrule
\end{tabular}
\end{table}

\begin{table*}[h!]
\centering
\caption{
Example of reasoning trace generated by Rank1~\citep{weller2025rank}, given a search query issued by the gpt-oss-20b agent and one retrieved passage for Query~37 on BrowseComp-Plus. 
The passage is irrelevant to Query~37.
The agent’s most recent reasoning trace is: ``$\cdots$We will search for ‘radiation’, ‘section 4.1’, ‘protein’, ‘two species’, and ‘2010’.''
Rank1 incorrectly treats the independent keywords ``radiation'' and ``protein'' as a coherent semantic unit, ultimately leading to an incorrect prediction.
}
\label{table_example_rank1}
\small
\setlength{\tabcolsep}{4pt} % optional: tighter columns
\renewcommand{\arraystretch}{1.15} % optional: a bit more row height

\begin{tabularx}{\linewidth}{>{\raggedright\arraybackslash}p{2.5cm}
                            >{\raggedright\arraybackslash}p{5cm}
                            >{\raggedright\arraybackslash}p{8.4cm}
                            >{\centering\arraybackslash}p{0.8cm}}
\toprule
Search query & Passage & Reasoning text & Score \\
\midrule
"section 4.1" radiation protein 2010 &
$\cdots$ A subset panel of significantly altered proteins was selected to build predictive models of radiation exposure and received radiation dose useful for population screening in a future radiological or nuclear event. $\cdots$ &
$\cdots$ The query mentions "section 4.1" which might refer to a specific section in a document or paper. The term "radiation protein" could be a typo or shorthand for a specific protein related to radiation exposure. $\cdots$ The query might be referencing a specific section in a paper titled "radiation protein 2010." Since the passage is from a 2010 study and discusses radiation-related proteins, it's likely relevant. $\cdots$ &
0.999 \\
\bottomrule
\end{tabularx}
\end{table*}

\vspace*{-2mm}
\subsection{Re-ranking in deep research}
\label{sec_rq2}
%\vspace*{-1.5mm}

To answer \ref{RQ2}, we evaluate two agents (gpt-oss-20b and GLM-4.7-Flash) that access ranking pipelines with varying retrievers, re-rankers, and re-ranking depths. 
Given the advantages of the passage corpus shown in Section~\ref{sec_rq1}, we perform all experiments on the passage corpus.
We use BM25, SPLADE-v3, and Qwen3-Embed-8B as initial retrievers; monoT5-3B, RankLLaMA-7B, and Rank1-7B as re-rankers; and, following \citep{sharifymoghaddam2026rerank}, we use three re-ranking depths (10, 20, and 50). 
Due to limited GPU resources, for GLM-4.7-Flash, we exclude Qwen3-Embed-8B, Rank1 and depth 20.
We show the results for gpt-oss-20b and GLM-4.7-Flash in Tables~\ref{tab_rerank_oss} and~\ref{tab_rerank_glm}, respectively.
We make three key observations. 

First, re-ranking consistently improves recall and accuracy while typically reducing search calls compared to no re-ranking. 
Notably, gpt-oss-20b with the BM25--monoT5 pipeline (depth 50) achieves the best performance (recall and accuracy) in our study, reaching 0.716 recall and 0.689 accuracy with 27.57 search calls. 
Relative to no re-ranking, this represents gains of 16.23\% in recall and 20.45\% in accuracy, alongside a 10.98\% reduction in search calls.
Despite using only a 20B agent with a BM25 retriever and a 3B re-ranker, this configuration achieves accuracy comparable to a GPT-5–based agent using Qwen3-Embed-8B (0.701; Table~1 in~\citep{chen2025browsecomp}).

Second, no single re-ranker consistently performs best.
Interestingly, the reasoning-based re-ranker Rank1~\citep{weller2025rank} does not show a clear advantage over non-reasoning ones.
Table~\ref{table_example_rank1} presents a failure case in which Rank1 misinterprets the search intent by incorrectly treating the independent keywords ``radiation'' and ``protein'' as a semantic unit, finally leading to a wrong relevance prediction.
This suggests that keyword-rich, phrase-driven web-search queries may reduce the effectiveness of Rank1’s explicit reasoning.
Rank1 is trained on MS MARCO~\citep{bajaj2018ms} that consists of natural-language questions, resulting in a training--inference query mismatch with agent-issued queries. 
We examine this issue in detail in Section~\ref{sec_rq3}.
%Rank1 struggles with keyword-style agent queries, reducing the effectiveness of Rank1’s reasoning.

%  and stronger initial retrievers 
Third, deeper re-ranking tends to improve effectiveness while reducing search calls.
For gpt-oss-20b with the BM25--monoT5 pipeline, increasing the depth from 10 to 20 improves recall and accuracy by 6.21\% and 6.81\%, respectively, while reducing search calls by 3.11\%. 
Further increasing the depth from 20 to 50 yields gains of 2.14\% in recall and 2.23\% in accuracy, alongside a 4.80\% reduction in search calls.
These trends align with~\citep{sharifymoghaddam2026rerank}, which reports improvements with deeper depths for listwise re-rankers in deep research.
%
%Regarding initial retrievers, with gpt-oss-20b and monoT5 (depth 50), BM25 as the retriever yields relative improvements of 16.61\% in recall and 23.26\% in accuracy over Qwen3-Embed-8B, while reducing search calls by 16.30\%.

%BM25 yields relative improvements of 3.92\% in recall and 6.66\% in accuracy over SPLADE-v3, while reducing search calls by 7.23\%; compared to Qwen3-Embed-8B, BM25 achieves larger relative gains of 16.61\% in recall and 23.26\% in accuracy, with a 16.30\% reduction in search calls.

\begin{table}[t!]
\centering
\caption{
Performance of the gpt-oss-20b agent under different retrievers and query conditions on the BrowseComp-Plus passage corpus.
Q2Q denotes our query-to-question method (Section~\ref{sec_q2q}); Q uses only the raw agent-issued query, while Q+R additionally incorporates the agent’s recent reasoning trace.
Within each retriever, the best values in recall and accuracy are \textbf{boldfaced}, and the second-best ones are \underline{underlined}.
$^*$ denotes a statistically significant improvement of Q2Q (Q+R) over raw agent-issued queries (paired $t$-test, $p < 0.05$).
}
\label{tab_q2q_oss}
\setlength{\tabcolsep}{3.2pt}
\renewcommand{\arraystretch}{1.05}
\begin{tabular}{l l cccc}
\toprule
Retriever & Query & \#Search & Recall & Acc. & Comp. \\
\midrule
\multirow{3}{*}{BM25}
& Raw        & 30.97 & \textbf{0.616} & \underline{0.572} & 0.968 \\
& Q2Q (Q)    & 31.88 & \underline{0.593} & \textbf{0.578} & 0.977 \\
& Q2Q (Q+R)  & 32.15 & 0.583 & 0.557 & 0.974 \\
\midrule
\multirow{3}{*}{SPLADE-v3}
& Raw        & 32.75 & 0.545 & \underline{0.516} & 0.980 \\
& Q2Q (Q)    & 32.88 & \underline{0.550} & 0.510 & 0.976 \\
& Q2Q (Q+R)  & 31.70 & \phantom{0}\textbf{0.585}$^*$ & \phantom{0}\textbf{0.557}$^*$ & 0.983 \\
\midrule
\multirow{3}{*}{\shortstack[l]{Qwen3\\-Embed-8B}}
& Raw        & 37.83 & \underline{0.470} & \underline{0.417} & 0.963 \\
& Q2Q (Q)    & 37.42 & 0.457 & 0.404 & 0.969 \\
& Q2Q (Q+R)  & 35.82 & \phantom{0}\textbf{0.507}$^*$ & \phantom{0}\textbf{0.459}$^*$ & 0.965 \\
\bottomrule
\end{tabular}
\end{table}

\begin{table}[t!]
\centering
\caption{
Examples of search queries issued by the gpt-oss-20b agent, and queries reformulated by the Q2Q method, for Query~781 on BrowseComp-Plus.
%Q2Q (Q) uses only the raw web search query generated by the agent, while Q2Q (Q+R) additionally incorporates the agent's recent reasoning trace.
The agent’s recent reasoning trace is:
``$\cdots$ Let's recall some matches that had an attendance of exactly 61,728, etc. Perhaps it may be easier to search for ‘attendance 61,880’ ‘football.’''
}
\label{tab_q2q_example}
\setlength{\tabcolsep}{4.5pt}
\begin{tabular}{l p{0.75\columnwidth}}
\toprule
Method & Search query \\
\midrule
Raw query  
& ``61,880'' football attendance \\
Q2Q (Q)  
& What is the football attendance number 61,880? \\
Q2Q (Q+R) 
& What football match had an attendance of 61,880? \\
\bottomrule
\end{tabular}
\end{table}

\begin{table}[t!]
\centering
\caption{
Performance of the gpt-oss-20b agent using a ranking pipeline with the SPLADE-v3 retriever and the Rank1 re-ranker on the BrowseComp-Plus passage corpus.
SPLADE-v3 always uses raw agent-issued queries, while Rank1 operates under different query conditions.
%
%The best values in recall and accuracy are \textbf{boldfaced}, and the second-best ones are \underline{underlined}.
%$^*$ denotes a statistically significant improvement of Q2Q (Q+R) over raw agent-issued queries (paired $t$-test, $p < 0.05$).
Metric definitions and formatting follow Table~\ref{tab_q2q_oss}.
}
\label{tab_oss_q2q_rerank}
\setlength{\tabcolsep}{3pt}
\renewcommand{\arraystretch}{0.8}

\begin{tabular}{l l cccc}
\toprule
Re-ranker & Query & \#Search & Recall & Acc. & Comp. \\
\midrule
\multirow{2}{*}{Rank1 ($d=10$)}
& Raw        & 31.18 & 0.617 & 0.580 & 0.978 \\
& Q2Q (Q+R)  & 31.15 & \phantom{0}\textbf{0.638}$^*$ & \phantom{0}\textbf{0.613}$^*$ & 0.970 \\
\bottomrule
\end{tabular}
\end{table}

%\subsection{Query mismatch between agent-issued and training queries}
%\subsection{Query mismatch in neural ranking}
\vspace*{-3mm}
\subsection{Training--inference query mismatch}
\vspace*{-0.5mm}
\label{sec_rq3}
% towards re-rankers

To address \ref{RQ3}, we evaluate the gpt-oss-20b agent using BM25, SPLADE-v3, and Qwen3-Embed-8B as retrievers under three search-query conditions: agent-issued queries, and questions generated by two variants of our query-to-question (Q2Q) method (see Section~\ref{sec_q2q}). 
%The Q variant uses only the raw query, while Q+R additionally incorporates the agent’s recent reasoning trace.
The two variants generate a question using only the raw query issued by the agent (denoted as Q) or using both the raw query and the agent's recent reasoning trace (denoted as Q+R).
Following Sections~\ref{sec_rq1} and~\ref{sec_rq2}, experiments are performed on the passage corpus. 
Due to limited GPU resources, we exclude GLM-4.7-Flash.
We present results in Table~\ref{tab_q2q_oss}.

We make three observations.
First, for both retrievers (SPLADE-v3 and Qwen3-Embed-8B), Q2Q (Q+R) consistently achieves significant improvements over raw agent-issued queries. 
E.g., with SPLADE-v3, Q2Q (Q+R) yields relative gains of 7.34\% in recall and 7.95\% in accuracy.
These results indicate that mismatch between agent-issued queries and the natural-language questions used to train neural rankers severely limits their effectiveness in deep research. 
Q2Q (Q+R) effectively mitigates this training–inference query mismatch.
Second, Q2Q (Q) provides little to no improvement over raw agent-issued queries.
To illustrate why, Table~\ref{tab_q2q_example} illustrates examples of a raw query generated by the gpt-oss-20b agent and its reformulations by Q2Q (Q) and Q2Q (Q+R).
The agent’s search intent is to retrieve football matches with an attendance of 61,880. 
The raw query (``61,880'' football attendance) is keyword-driven and under-specified. 
Q2Q (Q) reformulates it as ``What is the football attendance number 61,880?'', shifting the focus from identifying matches to explaining the number itself. 
In contrast, Q2Q (Q+R), which incorporates the agent’s reasoning trace, better captures the search intent.
Third, for BM25, Q2Q-generated questions even hurt performance, indicating that web-search-style queries are better suited to BM25 than natural-language questions.

Given that Q2Q (Q+R) reduces query mismatch for neural retrievers, we further investigate whether it also alleviates mismatch in re-ranking. 
Section~\ref{sec_rq2} shows that keyword- and phrase-based web-search queries can weaken the reasoning effectiveness of the re-ranker Rank1~\cite{weller2025rank}.
We therefore evaluate gpt-oss-20b using a SPLADE-v3--Rank1 pipeline (re-ranking depth $d=10$). 
SPLADE-v3 uses raw agent-issued queries, while Rank1 uses the raw queries or the Q2Q (Q+R) reformulations.
As shown in Table~\ref{tab_oss_q2q_rerank}, Q2Q (Q+R) significantly mitigates query mismatch for Rank1, yielding relative gains of 3.40\% in recall and 5.69\% in accuracy over raw queries.
%\vspace*{-4.8mm}
\section{Related Work}
%\vspace*{-1.5mm}

\header{Text ranking}
Unsupervised lexical retrievers (e.g., BM25~\citep{Robertson1994OkapiAT}) have long dominated text ranking. 
With the rise of pre-trained language models (e.g., BERT~\citep{devlin2019bert} and T5~\citep{raffel2020exploring}) and large-scale human-labelled training data~\citep{bajaj2018ms}, neural rankers have rapidly advanced.
A wide range of neural retrievers has been developed, including single-vector dense~\citep{xiong2021approximate}, multi-vector dense~\citep{santhanam2022colbertv2,khattab2020colbert}, and learned sparse retrievers~\citep{lassance2024splade,formal2021splade}.
In addition, cross-encoder re-rankers~\citep{nogueira2020document} based on BERT~\citep{devlin2019bert} or T5~\citep{raffel2020exploring} have achieved strong effectiveness in re-ranking.
More recently, LLMs have further advanced neural ranking~\citep{meng2026re}, e.g., through stronger representation capabilities~\citep{ma2024fine} and large-scale synthetic training data~\citep{zhang2025qwen3}. 
LLM reasoning has also been explored to enhance re-ranking~\citep{weller2025rank,yang2025rank,zhang2025entropy}.
Text ranking has been studied across diverse settings, including single-hop retrieval~\citep{bajaj2018ms,kwiatkowski2019natural}, multi-hop retrieval~\citep{ho2020constructing,yang2018hotpotqa}, reasoning-intensive search~\citep{shao2025reasonir,hongjinbright}, and RAG systems~\citep{mo2026opendecoder,su2025parametric,mo2025uniconv,asai2024selfrag}. 
However, little work has systematically examined the performance of text ranking methods in deep research.

\header{Deep research}
The origins of deep research can be traced back to multi-hop question answering (QA)~\citep{ho2020constructing,yang2018hotpotqa,trivedi-etal-2022-musique}. 
Most multi-hop QA benchmarks rely on Wikipedia, which has been extensively used in LLM pre-training, making such questions less challenging today~\citep{chen2025browsecomp}.
In contrast, deep research requires iterative web-scale search and evidence synthesis across the open web~\citep{wei2025browsecomp,zhou2025browsecomp}.
E.g., the BrowseComp benchmark~\citep{wei2025browsecomp} includes queries that humans typically cannot solve within ten minutes using search engines. 
Solving these queries typically requires LLM-based agents to perform multiple rounds of search and reasoning to gather, synthesise, and verify evidence~\citep{ou2025browseconfconfidenceguidedtesttimescaling,tongyideepresearchteam2025tongyideepresearchtechnicalreport}.
Most existing approaches equip LLM-based agents with live web search APIs~\citep{li2025websailor,wang2026deepresearcheval}, but such black-box systems lack transparency.
To address this issue, \citet{chen2025browsecomp} curate BrowseComp-Plus, which extends the original BrowseComp dataset with a fixed document corpus and human-verified relevance judgments, allowing white-box retrievers to be used.
On this resource, \citet{sharifymoghaddam2026rerank} study listwise LLM-based re-rankers to analyse effectiveness–efficiency trade-offs.
We differ from prior studies~\citep{chen2025browsecomp,sharifymoghaddam2026rerank} by systematically reproducing a broad spectrum of text ranking methods in deep research. 
Unlike contemporaneous work~\citep{hu2026sage} that tests BM25 and LLM-based single-vector retrievers in scientific literature search (domain-specific deep research), we use a broader range of retrievers (e.g., learned sparse and multi-vector) and re-rankers in a general open-domain setting.

%, hindering systematic analysis and clear understanding of the contribution of search components towards solving deep research queries.
%
%

%Muti hope QA
%Web search 
%harder,
%not loacal knowledge base, web search
%using  search APIs \citep{xu2026self}

%\citep{xia2025open} propose a framework to synthesise deep research training data.
%\citep{zhang2025memory} formulating context management as in-place editing operations (deletion, insertion) context management
%
%\citep{yu2025browseragent} propose a agent that can perform a set of predefined browser actions (e.g., scroll up and down) on raw web pages.
%
%\citep{luo2025infoflow} focus on mitigating the issue of low reward density over iterations of reasoning and searches.

%\vspace*{-3.0mm}
\section{Conclusions \& Future Work}
%\vspace*{-1mm}

We have reproduced an extensive set of text ranking methods in deep research, conducting experiments on BrowseComp-Plus~\citep{chen2025browsecomp} with 2 open-source agents, 5 retrievers, and 3 re-rankers.

Overall, our results show that text ranking methods developed by the IR community are still highly effective in deep research.
Several well-established findings continue to hold:
\begin{enumerate*}[label=(\roman*)]
\item the lexical retriever BM25 with appropriate setup outperforms neural rankers in most cases; notably, gpt-oss-20b with BM25 on the passage corpus achieves the highest answer accuracy across all retrieval settings in our study;
\item BERT-based learned sparse~\citep{lassance2024splade} and multi-vector dense~\citep{santhanam2022colbertv2} retrievers generalise better than LLM-based single-vector dense retrievers; and
\item re-ranking remains highly effective, improving recall and answer accuracy while reducing search calls, with deeper re-ranking depths further amplifying these gains.
\end{enumerate*}

However, the web-search-style syntax of agent-issued queries (e.g., quoted exact matches) induces distribution drift for neural rankers, limiting their effectiveness in deep research and preventing a prior finding from generalising to this setting: the reasoning-based re-ranker Rank1~\citep{weller2025rank} often misinterprets such queries, diminishing the benefits of reasoning and showing no clear advantage over non-reasoning methods.
Our proposed Q2Q method significantly mitigates this drift, improving neural ranking performance.

Beyond validating these findings, we find that passage-level units benefit agents with limited context windows, and reduce sensitivity to document-length normalisation in lexical retrieval.

We identify future directions: 
\begin{enumerate*}[label=(\roman*)]
%\item it is worthwhile to augment other deep research datasets with a fixed corpus and document judgements;
\item we only use two LLMs as agents; evaluating additional model families and larger model sizes would help assess the generalisability of our findings;
\item exploring additional rankers~\citep{akram2026jina,shao2025reasonir,li2023towards} and configurations, such as scaling laws, is an important direction; and
\item extending deep research to conversational scenarios~\citep{meng2023system,mo2025uniconv,meng2025bridging,meng2026conversational} is a promising avenue.
\end{enumerate*}

%Passage-level units enable more search and reasoning iterations before agents reach context limits, particularly benefiting agents with shorter context windows, while reducing sensitivity to document-length normalisation in lexical retrieval.
%We show that findings on traditional text ranking partially generalise to this new setting. 
%We find that neural retrievers often still underperform the lexical sparse retrievers in the deep research setup. 
%Re-ranking effectively improves agents’ retrieval performance and answer accuracy; however, Rank1, the reasoning-based re-ranker, does not show a clear advantage over non-reasoning-based methods because it struggles with keyword-style queries issued by agents, which reduces the effectiveness of its reasoning.
%
%Passage-level retrieval yields higher end-to-end answer accuracy than document-level retrieval, consistent with enabling more search/reasoning iterations before agents hit context or output limits.
%
%Lexical retrieval is particularly strong in this setting: agent queries often resemble web-search syntax (including quoted exact matches), aligning well with BM25-style matching.

%We systematically revisit text ranking for deep research agents on BrowseComp-Plus and find that retrieval design choices shift when ranking is embedded inside iterative agent loops. 
% A natural next step is to train/adapt rankers on agent-generated query distributions and interaction traces to reduce mismatch without relying on external rewriting.  

%In this work, we have 
%only one dataset
%we only consiser limited browsing tools
\begin{acks}
We thank Zijian Chen and Xueguang Ma for their guidance on using BrowseComp-Plus.
This research was supported by a Turing AI Acceleration Fellowship funded by the Engineering and Physical Sciences Research Council (EPSRC), grant number EP/V025708/1.
\end{acks}

%\clearpage
%\input{sections/08_acknowledgement}
%\clearpage
\bibliographystyle{ACM-Reference-Format}
\balance
\bibliography{references}

\end{document}